\documentclass[twocolumn,showpacs,amssymb,prd]{revtex4}

\usepackage{graphicx}
\usepackage{dcolumn}
\usepackage{bm}
\usepackage{epsfig}

\begin{document}
\newcommand{\gsim}{\hbox{\rlap{$^>$}$_\sim$}}
\newcommand{\lsim}{\hbox{\rlap{$^<$}$_\sim$}}

\title{Universal Afterglow Of Supernova-Less Gamma Ray Bursts}

\author{Shlomo Dado and Arnon Dar}
\affiliation{Physics Department, Technion, Haifa, Israel}
      
\begin{abstract} 
The well-sampled early time afterglows of gamma ray bursts (GRBs) not 
associated with a supernova (SN) explosion, can be scaled down to a simple 
dimensionless universal formula, which describes well their temporal behavior. 
Such SN-less GRBs include short hard bursts  and long bursts
The universal behavior of their afterglows is that 
expected from a pulsar wind nebula powered by the rotational energy loss 
of the newly born millisecond pulsar.

\end{abstract}

\pacs{98.70.Sa,97.60.Gb,98.20}

\maketitle

\noindent 
{\bf Introduction.} Gamma-ray bursts (GRBs) are brief flashes of gamma rays 
lasting between few milliseconds and several hours [1] from extremely 
energetic cosmic explosions [2]. They were first detected in 1967 by the USA 
Vela satellites. Their discovery was published in 1973 after 15 such events 
were detected [3]. GRBs fall roughly into two classes [4], long duration ones 
(LGRBs) that last more than $\sim$ 2 seconds, and short hard bursts (SGRBs) 
that typically last less than 2 seconds. For 3 decades after their discovery, 
the origin of both types was completely unknown. This has changed 
dramatically by the first X-ray localization of GRBs and the discovery of 
their X-ray afterglow with the BeppoSAX satellite, which led also to the 
discovery of GRBs' afterglow at longer wave lengths, their host galaxies and 
their redshifts [5], and to the detailed measurements with ground and space 
based telescopes of the properties of their prompt and afterglow emissions, 
their host galaxies, and their near environments.
 
The late-time afterglow of GRB970228, the first localized GRB by 
BeppoSAX, also included photometric evidence of an associated supernova 
[6] which met skepticism, as did [7] the original suggestions of a 
GRB-SN association [8] long before this first observational evidence. 
Only when photometric and spectroscopic evidence [9] for other SN-LGRB 
associations has been accumulated from relatively 
nearby LGRBs, the SN-LGRB association became widely accepted. Moreover, 
it was also believed (e.g. [10]) that in all ordinary LGRBs where an 
associated supernova of type Ic akin to 1998bw was not seen, it was 
because it was too distant, and/or overshined by the GRB afterglow 
and/or by the light of the host galaxy, or simply was not looked for.

However, deep optical searches of SNe associated with several relatively 
nearby LGRBs have failed to detect an associated SN [11]. They provided 
compelling evidence that SN explosions are not the only source of LGRBs. 
But their origin has not been established beyond doubt and is still 
debated.

As for SGRBs, until recently they were widely believed to be produced in 
neutron stars mergers (NSMs) in compact binaries [12], as first 
suggested three decades ago [13], and perhaps in neutron star-black hole 
mergers [14]. This wide belief was based on indirect evidence [12]. 
Recently, however, the short hard burst SHB170817A that followed $\sim$ 
1.7 s after GW170817 [15], the first direct detection of gravitational 
waves (GWs) emitted from neutron stars merger, by the Ligo-Virgo GW 
detectors [16], has shown beyond doubt that neutron star mergers produce 
SGRBs.

In this letter, we show that all the well sampled X-ray afterglows of 
SN-less GRBs within the first couple of days after burst have a temporal 
behavior, which can be scaled down to a simple dimensionless universal 
form. This universal form is expected if the afterglows of SGRBs and SN-less 
LGRBs during the first couple of days after burst are dominated by the 
emission of a pulsar wind nebula (PWN) powered by the newly born 
pulsar [17]. Several implications of this observations are shortly discussed.
 
\noindent {\bf Universal Afterglows.} 
As long as the spin-down of a pulsar 
with a period $P(t)$
satisfies $\dot{P}P \!=\!const $,  
\begin{equation} 
P(t)=P_i(1 +t/t_b)^{1/2}, 
\end{equation} 
where $P_i=P(0)$ is the initial period of the pulsar, $t$ is the 
time after its birth, and  $t_b=P_i/2\,\dot{P}_i$. 
If a constant fraction $\eta$ of the rotational energy loss of such pulsar 
is reradiated by the PWN, then, in a steady state, 
its luminosity  satisfies $L\!=\!\eta\, I\,w\, \dot{w}$, 
where $w=2\,\pi/P$ and $I$ is the moment of inertia of the neutron star.
Hence, in a steady state, the luminosity emitted by a PWN satisfies 
\begin{equation}
L(t)\!=\!L(0)/(1\!+\!t/t_b)^2.
\end{equation}
Eq.(2) can be written as
\begin{equation}
L(t)/L(0)\!=\!1/(1\!+\!t_s)^2, 
\end{equation}
where $t_s\!=\!t/t_b$.
Thus, the dimensionless luminosity $L(t)/L(0)$
has a simple universal form as function of 
the scaled time $t_s$. For each  
afterglow of an SN-less GRB powered by 
a pulsar, $L(0)$ and $t_b$ can be obtained from a best fit 
of Eq.(2) to the light curve of their measured afterglow. 

The initial period of the pulsar enshrouded within a PWN can be 
estimated from its locally measured energy flux $F(0)$ 
corrected for absorption along the line of sight to the PWN,
its redshift $z$, and its luminosity distance $D_L$.
\begin{equation}
P_i\!=\!{1\over D_L}\,\sqrt{{(1\!+\!z)\,\eta\, \pi\,I \over  2\,F(0)\,t_b}},
\end{equation}
where $F=L/4\,\pi\, D_L^2$,
$I\!\approx\!(2/5)\,M\,R^2\!\approx\! 1.12\times 10^{45}\,{\rm g\,cm^2}$, 
for a canonical pulsar with  $R\!=\!10$ km and $M\approx 1.4\, M_\odot $,
and $\eta\!<\!1$.  
The period derivative can be obtained from the relation 
$\dot{P}_i\!=\!P_i/2\,t_b$.

\noindent
{\bf Comparison with observations}.
In [18], we have fitted the X-ray lightcurves of all SGRBs with a well 
sampled afterglow measured with the Swift XRT [19], assuming the 
cannonball model for the prompt and extended emission and Eq.(2) for the
taking-over afterglow. Figure 1 demonstrates such a fit for SHB150424A.  
\begin{figure}[]
\centering
\epsfig{file=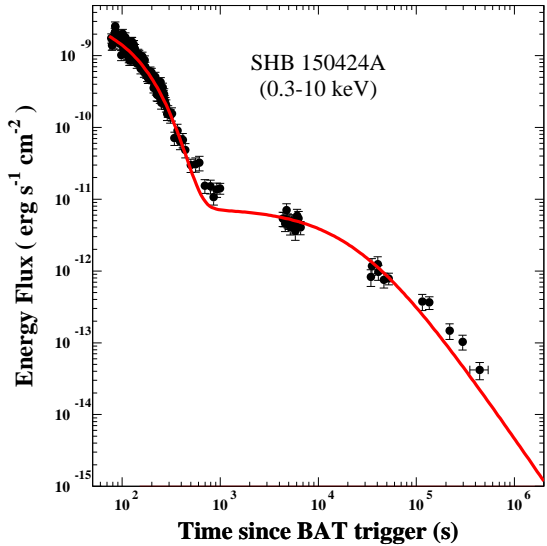,width=8.0cm,height=8.0cm}
\caption{Comparison between the observed X-ray afterglow  of SHB150424A
measured with Swift XRT [19] and the best fit assuming the cannonball model 
for the prompt and extended emission [18] and Eq.(2) for their late time 
afterglow.}
\end{figure}
Eqs,(2),(3) are expected to be valid only after the last accretion 
episode on the newly born pulsar, and after the PWN emission powered by 
the pulsar's power supply has reached a steady state. Since the exact 
times of both are not known, and in order to avoid a contribution from 
the prompt emission, we have fitted the observed afterglows of 
SN-less GRBs with Eq.(2) after the plateau phase took over the fast decline 
of the prompt emission (last pulse/flare or extended emission). This probably 
made unimportant the lack of knowledge of the exact birth time/power supply of 
the pulsar and when the PWN emission powered by
the pulsar's power supply has reached a steady state.
In Figure 2 we plotted the dimensionless X-ray afterglow of 12 
SGRBs with the best sampled afterglows measured with the Swift XRT [20] 
in the first couple of days after burst and the universal behavior given 
by Eq.(3). The values of $L(0)$ and $t_b$, needed to reduce each measured 
lightcurve to the dimensionless form, were obtained for each SGRB, from a 
best fit of Eq.(2) to the observed plateau after the fast decline 
phase of the X-ray afterglow.
\begin{figure}[]
\centering
\epsfig{file=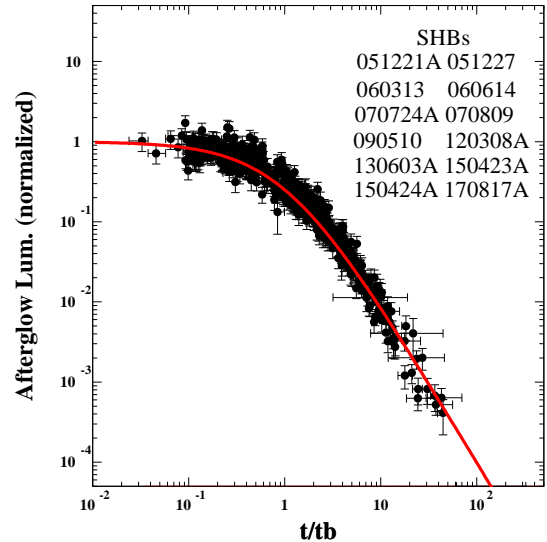,width=8.0cm,height=8.0cm}
\caption{Comparison between the dimensionless  light curve
of the X-ray afterglow of 12 SGRBs 
with a well sampled afterglow, measured with the Swift XRT [19]
during the first couple of days after burst 
and the predicted universal behavior given by Eq.(3).
The bolometric light curve of SHB170817A reported in [20]
is also included.} 
\end{figure}
The 

The predicted universal behavior of the afterglow of SN-less 
LGRBs as given by Eq.(2) is compared in Figure 3 to
the observed afterglow lightcurves of the long duration GRB990510,
in the X-ray [21] and optical I, R, V, B  bands,
[22] rescaled according to Eq.(3)  
and plotted as a function of $t/t_b$. 

\begin{figure}[]
\centering
\epsfig{file=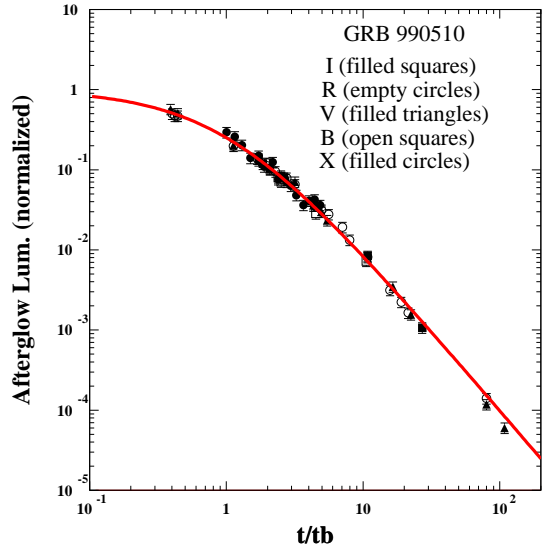 ,width=8.0cm,height=8.0cm}
\caption{Comparison between the observed afterglow of GRB990510 
in the X-ray [21] and optical I, R, V, B  bands 
[22] rescaled according to Eq.(2) and plotted as function of $t/t_b$, 
and their predicted universal form as given by Eq.(3).}
\end{figure}

Unlike SGRBs, LGRB seem to be divided to two classes [23],
SN-LGRBs and SN-less LGRBs. Only in relatively nearby long GRBs 
the GRB class can be identified by deep searches of an associated 
SN. However, SN-LGRBs are produced mostly in star formation 
regions within molecular clouds of relatively high density. 
In such cases the GRB afterglow seems to be dominated by 
the synchrotron radiation emitted from the deceleration of the  
highly relativistic jet in the dense ISM. The spectral energy density 
of the emitted afterglow is well described by a smoothly broken 
power-law with a spectral index $\beta$ and a temporal decay index
$\alpha$ which well after the "break" satisfies the closure relation
$\alpha\!=\!\beta+1/2$ predicted by the CB model of SN-LGRBs [24]. This 
relation seems to be well satisfied by SN-LGRBs but not by 
SN-less LGRBs. This was used to identify long SN-LGRBs whose afterglow
was produced by synchrotron emission from a decelerating 
highly relativistic jet [25].

The well sampled afterglows of a representative set of SN-less LGRBs that 
were measured with the Swift XRT during the first few days after burst, are 
compared in Figure 4 to their expected universal behavior as given by 
Eq.(3). The parameters $L(0)$ and $t_b$ needed to reduce each measured 
lightcurve to the dimensionless universal form were obtained from a best 
fit of Eq.(2) to the plateau followed by the fast decline phase of the 
X-ray afterglow of each GRB. 
\begin{figure}[]
\centering 
\epsfig{file=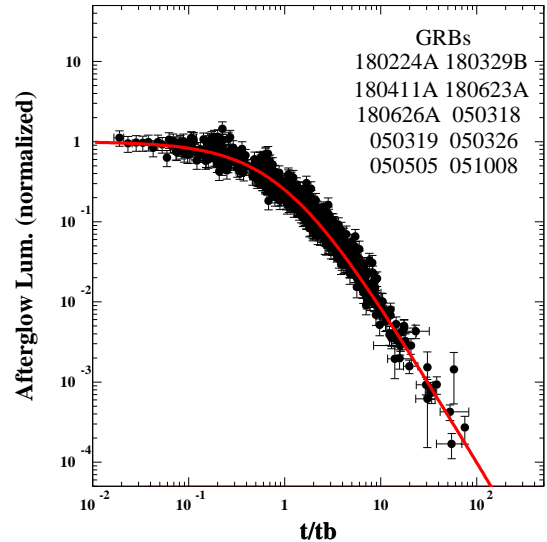,width=8.0cm,height=8.0cm} 
\caption{Comparison between the normalized light curve of the X-ray 
afterglow of 10 SN-less GRBs with a well sampled universal
X-ray lightcurve measured in 2005 and 2006
with Swift XRT [19] in the first couple of days after burst and their 
predicted universal behavior as given by Eq.(3). $\chi^2/dof\!=\!825/824$.} 
\end{figure} 

\begin{figure}[]
\centering
\vspace{2cm}
\vbox{
\hbox{
\epsfig{file=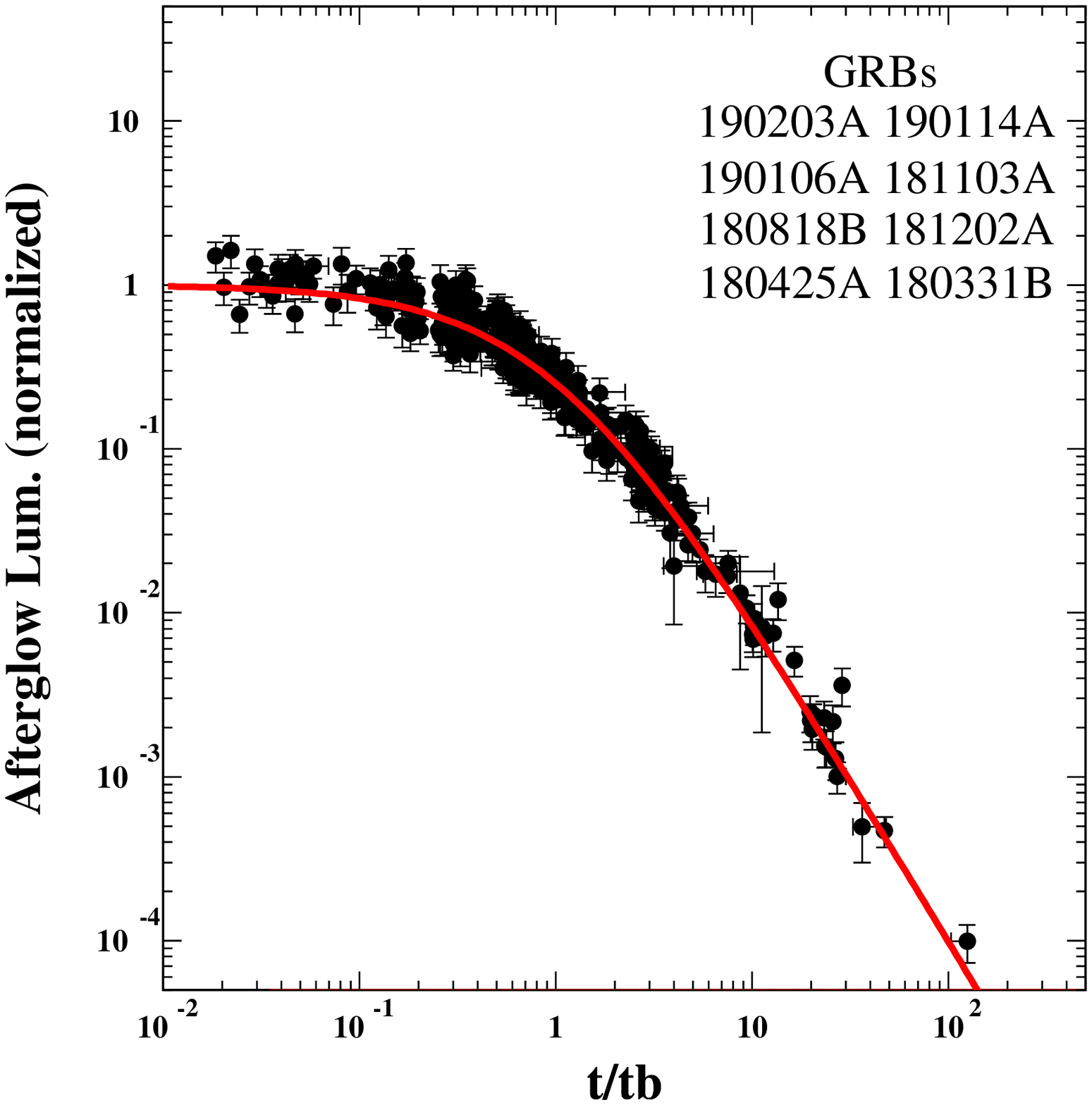,width=8.6cm,height=8.6cm}
\epsfig{file=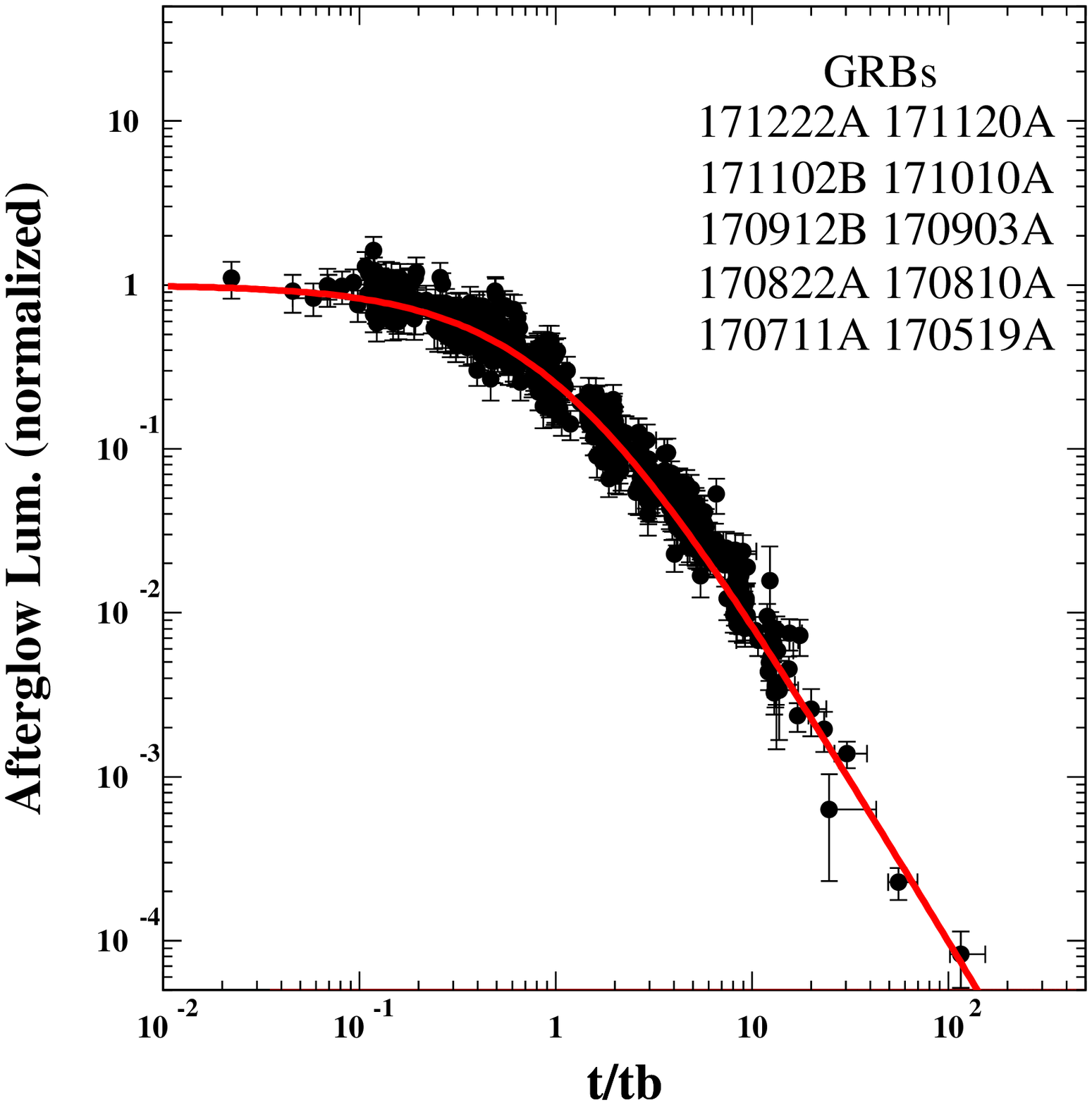,width=8.6cm,height=8.6cm}
}}
\vbox{
\hbox{
\epsfig{file=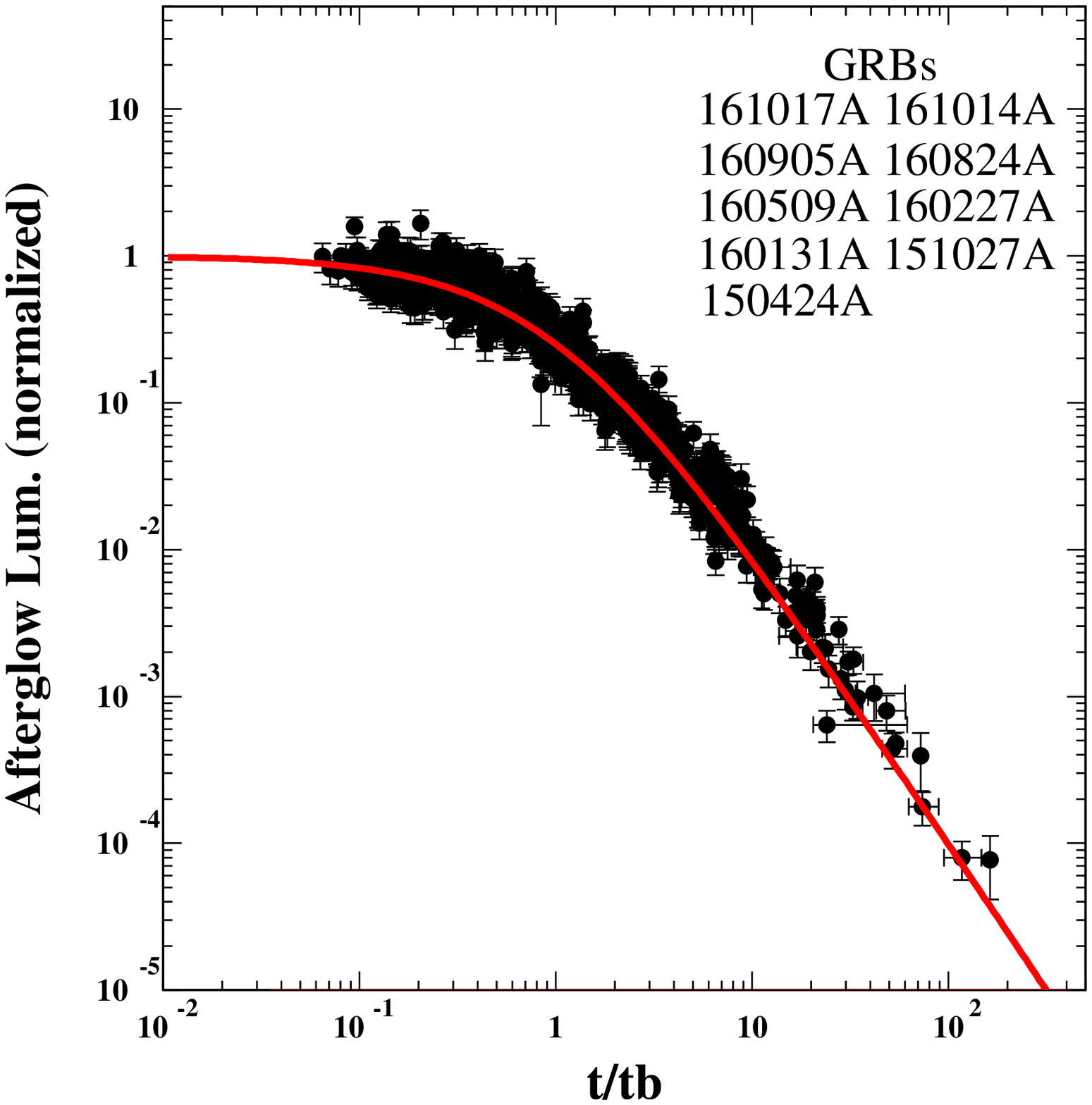,width=8.6cm,height=8.6cm}
\epsfig{file=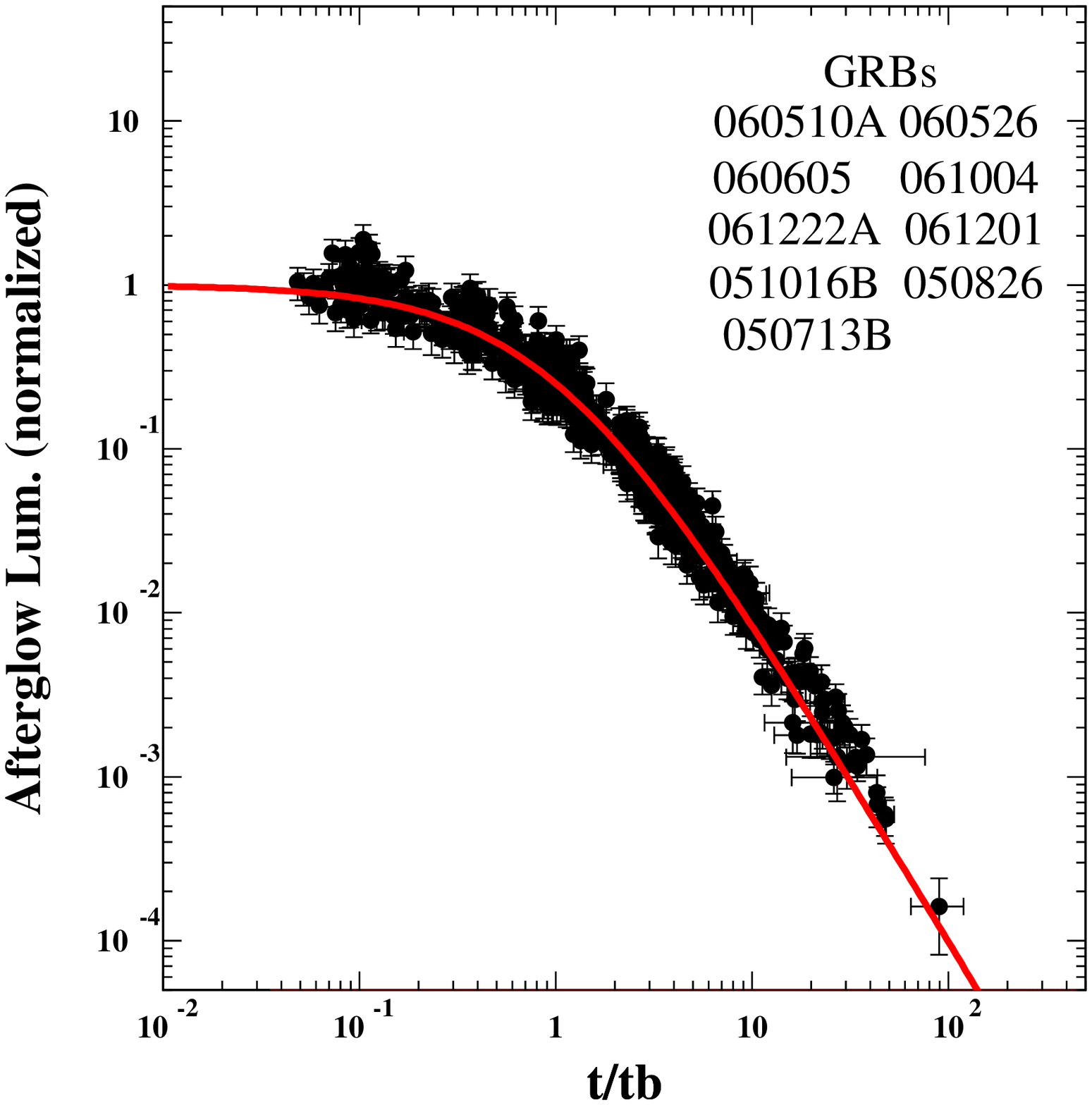,width=8.6cm,height=8.6cm}
}}
\caption{Comparison between the normalized light curve of the X-ray 
afterglow of SN-less GRBs with a well sampled universal
X-ray lightcurves measured with the Swift XRT [19] in the first couple of 
days after burst and their predicted universal behavior as given by Eq.(3). 
Topleft: $\chi^2/dof=353/345=1.02$. Top right:  $\chi^2/dof\!=\!718/528\!=
\!1.36$. Bottom left: $\chi^2/dof\!=\!2388/1588\!=\!1.48$ . Bottom right: 
$\chi^2/dof\!=\!845/747\!=\!1.13$.}  
\end{figure} 

Tables I,II  summarize the 
parameters in Eq.(2) which best reproduce    
the X-ray afterglow of the 
24 SN-less GRBs listed in Figures 1-4, 
and the pulsar periods derived from them
using Eq.(4) and assuming $\eta\!=\!1$.

\begin{table*}
\caption{}
\begin{tabular}{l l l l l l}
\hline
\hline
~~SHB~~~& ~~z~~~ & ~$F_X(0)$~ & ~$t_b$ & $\chi^2/dof$ &  $P$  \\
        &      &$[erg/s\,cm^2\,]$& ~[s]~&             &  [ms] \\
\hline
051221A &0.5465& 2.33E-12 &46276 &~~1.04 & 16.2  \\
051227  &0.8   & 1.44E-11 &~2280 &~~0.93 & 19.8  \\
060313  &      & 2.91E-11 &~4482 &~~0.60 &       \\
060614  &0.125 & 1.11E-11 &48931 &~~1.39 & 33.2  \\
070724A &0.457 & 1.18E-12 &18041 &~~1.46 & 43.7  \\
070809 &0.2187 & 3.231E-12&17181 &~~1.05 & 57.6  \\
090510  &0.903~& 3.68E-10 &~~551 &~~1.76 & 7.04  \\
120308A &      & 6.84E-11 &~4742 &~~1.51 &       \\
130603B &0.3564& 6.25E-12 &34382 &~~1.19 & 17.8  \\
150423A &1.394~& 1.31E-11 &~1290 &~~1.36 & 16.0  \\
150424A &0.30  & 5.15E-12 &34461 &~~1.51 & 23.5  \\
170817  &0.0093&          &117374&~~0.60 &       \\
\hline
\end{tabular}  
\end{table*}  

\begin{table*}
\caption{}
\begin{tabular}{l l l l l l}
\hline
~~GRB~~~& ~~z~~~ & ~$F_X(0)$~ & ~$t_b$ & $\chi^2/dof$ &  $P$  \\
        &      &$[erg/s\,cm^2\,]$& ~[s]~&             &  [ms] \\
\hline
050318  & 1.44 & 1.03E-10 &~3972 &~~0.92 & 3.14 \\
050319  & 3.24 & 1.85E-11 &27285 &~~1.45 & 1.39 \\
050326  &      & 1.50E-10 &~3138 &~~1.38 &      \\
050505  & 4.27 & 3.56E-11 &14059 &~~0.89 & 1.12 \\
051008  & 2.77 & 6.03E-11 &~5865 &~~1.41 & 1.88 \\
180224A &      & 1.02E-10 &~1537 &~~0.84 &      \\
180329B & 1.998& 2.54E-11 &~8542 &~~0.94 & 2.15 \\
180411A &      & 9.30E-11 &~9510 &~~0.98 &      \\
180623A &      & 2.17E-10 &~2498 &~~0.93 &      \\
180626A &      & 2.70E-11 &12110 &~~1.18 &      \\
\hline
\end{tabular}
\end{table*}
We have verified that almost all the hundreds LGRBs with a well sampled 
X-ray afterglow in the first couple of days after burst that have been
measured since 2005 with the Swift XRT and show an initial plateau, but 
are without an identified SN association and do not satisfy the late 
time CB model closure relation of SN-LGRBs, 
seem to satisfy well Eqs.(2),(3) ($\chi^2/dof\!\sim\! 1$).
This is demonstrated in Figure 5 for a sample of  
40 LGRBs measured with SWIFT in the past 5 years.

{\bf Discussion and conclusions:}
{\it All} the well sampled afterglows of SGRBs within a few 
days after burst are well described by Eqs.(2),(3) as shown in 
Figures 1,2. That, and the detection of SHB170817A [15], which 
followed the detection of gravitational waves (GWs)  from the 
GW170817 by the Virgo-Ligo GW detector [16], 
indicates that SGRBs are produced mainly by neutron stars merger 
[13] and not by neutron star - black hole merger [14]
in compact binaries.

LGRBs seem to consist of two distinct populations, SN-less 
LGRBs and SN-LGRBs. An SN-less identity was established 
observationally only for relatively nearby LGRBs by very deep 
searches [11]. In more distant LGRBs, an SN-less identity could 
not be established because the SN could have been overshined 
by the GRB afterglow and/or the host galaxy, or simply was not 
looked for.

An indirect way of identifying SN-less LGRBs  is the characteristic 
universal afterglow of SN-less LGRBs, which is very different from 
the afterglow of of SN-LGRBs: The late-time afterglow of SN-LGRBs 
seems to have a spectral energy flux density, which is well described 
by $F_\nu\!\propto\! t^{-\alpha}\, \nu^{-\beta}$ with
$\alpha \!=\!\beta\!+\!1/2$, as predicted [24] by the cannonball 
(CB) model of SN-LGRBs, where a highly relativistic jet of CBs ejected  
in an SNIc explosion produces the afterglow by synchrotron radiation 
emitted during their deceleration in the interstellar medium (ISM). 

The afterglow of SN-less LGRBs seems to be quite different. It has a 
simple temporal behavior  well described by Eq.(2) and can 
be scaled down to the universal  behavior given by 
Eq.(3). This was demonstrated in Figure 3 for GRB990510 and in Figure 
4, for clarity, only for all LGRBs with well sampled afterglow 
measured with the Swift XRT during its first year and last year 
of observations.

Figure 3 also demonstrates that the 
claims in the title of [21] "BeppoSAX confirmation of beamed 
afterglow emission from GRB 990510" and of [22] "Optical and 
Radio Observations of the Afterglow from GRB990510: Evidence for 
a Jet" which were based on arbitrary parametrizations and
beliefs rather than critical science were misleading.

Figure 4 clearly demonstrates that although the relation 
$\dot{P}\,P \!=\!const$ has been derived [26] for pulsars which 
spin down by magnetic dipole radiation (MDR) (in vacuum, 
assuming time-independent magnetic field and moment of inertia 
during spin down), the GRB data suggests that it may be more 
general. For instance, it is satisfied to a good accuracy by the 
Crab pulsar, despite the fact that the total luminosity of the 
Crab PWN which is powered by the Crab pulsar, is much higher 
than its MDR luminosity, estimated from its current $P$ and 
$\dot{P}$. This power supply to the PWN can be by pulsar
cosmic ray particles and relativistic winds.

Indeed, relativistic wind (RW) particles and high energy cosmic 
rays (CRs)  with $E\!\approx\!p\,c$, which spiral out along the 
open magnetic field lines and escape at the light 
cylinder (of a radius $c/w$ around the rotation axis) carry out energy 
and angular momentum at a rate $\dot{E}= Lum(CR)\!+\!Lum(RW)$ and 
$\dot{l}\!=I\, \dot{w}\!=\!Lum/ w$, respectively, where 
$Lum\!=\!Lum(CR)\!+\!Lum(RW)$ is the energy loss rate by CRs and RW. 
If this loss of angular momentum dominates the spin down of a pulsar, then 
the estimate [26] of its magnetic field at the magnetic poles,
$B_p\!=\!6.4\times 10^{19} \sqrt{P\, \dot{P}}$ Gauss,
is an over estimate. Moreover, this estimate assumes
a vacuum environment and a time-independent 
magnetic field [26] of the newly born milli second pulsar (MSP), and
cannot be trusted as solid evidence 
that MSPs which  seem to  power the 
afterglows of SN-less LGRBs  are magnetars [27].  
  
If  the afterglow of SN-less GRBs is powered by a newly born MSP, 
then the period which is obtained from best fits of Eq.(2) to
its  afterglow must yield  periods well above the classical 
Newtonian lower limit $P\!>\!2\pi \,R/c\!\approx\! 0.2 $ ms for canonical 
neutron stars. So far, all the fitted  SN-less GRBs have
yielded much larger periods than 0.2 ms.

It is however, quite remarkable that the inferred  periods of pulsars
in SGRBs are typically order of magnitude larger than those inferred  
in  SN-less LGRBs. That may be due to different $\eta$ values or loss 
of angular momentum by gravitational wave emission in neutron stars   
mergers compared to the spin up of neutron stars  in 
high mass X-ray binaries (HMXB) by mass 
accretion or phase transition. It may indicate that core collapse supernova 
explosions of massive stars are driven mainly by transport out of 
excessive rotational energy from infalling layers on the proto neutron star, 
which cannot be spun up to a surface velocity exceeding the speed of light light. 
This may be even more important than energy momentum deposition by 
neutrinos and shock waves in the 
envelope of the star, which so far, in numerical simulations, could 
not produce consistently core collapse SN explosions of massive stars 
with  kinetic energy  $E_k\!\sim\! 10^{51}$ erg [29].

\noindent
{\bf Acknoweledgement} The authors would like to thank Erez Ribak for a very 
useful suggestion and comments.

\end{document}